\documentclass[twocolumn,preprintnumbers,amsmath,amsfonts,amssymb,notitlepage,showpacs,aps,prb,longbibliography]{revtex4-2}

\usepackage{color}

\date{\today}

\usepackage{amsmath}
\DeclareMathOperator{\Tr}{Tr}

\usepackage{graphicx}

\usepackage{hyperref}

\def \be{\begin{equation}}
\def \ee{\end{equation}}
\def \ba{\begin{array}}
\def \ea{\end{array}}
\def \bea{\begin{eqnarray}}
\def \eea{\end{eqnarray}}

\newcommand{\VZZ}{\begin{smallmatrix}Z \\Z \end{smallmatrix}}
\newcommand{\HZZ}{\begin{smallmatrix} Z\, Z \end{smallmatrix}}
\newcommand{\VXX}{\begin{smallmatrix}X \\X \end{smallmatrix}}
\newcommand{\HXX}{\begin{smallmatrix} X\, X \end{smallmatrix}}

\begin{document}
\title{Subsystem symmetry,
spin glass order, and criticality  from random measurements in a two-dimensional Bacon-Shor circuit
}
\author{Vaibhav Sharma}
\email{vs492@cornell.edu}
\author{Chao-Ming Jian}
\email{chao-ming.jian@cornell.edu}
\author{Erich J Mueller}
\email{em256@cornell.edu}
\affiliation{Laboratory of Atomic and Solid State Physics, Cornell University, Ithaca, New York}

\begin{abstract}

We study a 2D measurement-only random circuit  motivated by the Bacon-Shor error correcting code. We 
find a rich phase diagram  as one varies the relative probabilities of
measuring nearest neighbor Pauli XX and ZZ check operators.
In the Bacon-Shor code, these checks commute with a group of stabilizer and logical operators, which therefore represent conserved quantities.  Described as a subsystem symmetry, these  conservation laws
lead to  a continuous phase transition between an $X$-basis and $Z$-basis spin glass order. The two phases are separated by a critical point 
where the entanglement entropy between two halves of an $L\times L$ system scales as $L\ln L$, a logarithmic violation of the area law. We generalize to a model where the check operators break the subsystem symmetries (and the Bacon-Shor code structure). In tension with established heuristics, we find that
the phase transition is replaced by a smooth crossover, and the $X$- and $Z$-basis spin glass orders spatially coexist. Additionally, if we approach the line of subsystem symmetries away from the critical point in the phase diagram, some spin glass order parameters jump discontinuously.     
\end{abstract}

\maketitle

\section{Introduction}\label{intro}

Random circuits provide a platform for exploring quantum dynamics and  quantum algorithms, revealing how information propagates and how entanglement evolves \cite{randcircuit1,randcircuit2,randcircuit3,randcircuit4,randcircuit5}. 
They provide simple examples of quantum dynamical systems ~\cite{quantumclassical,ChoiBaoQiEhud2020MIPT} and also act as models of noisy hardware \cite{NISQ}. A central insight has been the establishment of analogies between thermodynamic phase transitions and the behavior of quantum circuits consisting of both random unitary operations and measurements \cite{hybridcircuit1d,hybridcircuitextended,hybridcircuit2, Barkeshli1D,topoqcode,ChoiBaoQiEhud2020MIPT,JianYouVassuerLudwig2020MIPT,Guallans2020ScalableProbes,GullansHuse2020Purification,review,PotterVasseur2021EntanglementDynamics,measurementphases,Barkeshli2d, ChoiBaoQiEhud2020MIPT,hybrid2d,TangZhu2020,Turkeshi2022GeneralD}.
As one varies model parameters, such as the probabilities of various gates or measurements, 
these quantum circuits can exhibit different phases characterized by distinct entanglement scalings, which are separated from each other by continuous phase transitions. As exemplified in Refs. \onlinecite{ lavasani, sriram,Huse}, quantum circuits consisting of only measurements can also exhibit rich phase diagrams when the measurements performed at different time steps do not commute with each other. Apart from certain examples \cite{ lavasani, sriram, measurementphases,Barkeshli2d, Hastings2021dynamically}, the dynamical systems evolving under measurement-only quantum circuits 
have not been systematically explored
beyond one dimension. In this paper, we focus on a two-dimensional (2D) measurement-only circuit on a square lattice where we are able to explore the role of both global symmetries and “subsystem” symmetries in quantum dynamics. We analyze the phase transitions in this circuit using a detailed analysis of order parameters and correlation functions. We find that our 2D model displays intriguing behaviors that are not seen in 1D
measurement-only circuits. 

 Our model is motivated by the Bacon-Shor code \cite{baconshor}, which is a quantum error correcting code on a 2D square lattice that relies upon repeated measurement of {\em check} operators.  Also known as {\em gauge} operators, these are a set of non-commuting local operators which can be used to detect errors without scrambling the stored quantum information.  In the Bacon-Shor code, the check operators consist of the product of any two nearest neighbor 
 Pauli $X$ operators on a horizontal bond, 
and the product of any two nearest neighbor Pauli $Z$ operators
 on a vertical bond (Fig.~\ref{parameterspace} (a)).  We will denote these check operators as $\HXX$ and $\VZZ$.  While the check operators do not commute with one another, they do commute with the Bacon-Shor stabilizer group, generated by the products of $Z$ operators along any two horizontal rows, and the products of $X$ operators along any two vertical columns (Fig.~\ref{parameterspace} (b),(c)).  Both the Bacon-Shor stabilizers and the check operators commute with logical operators:  A product of $Z$'s along a single row or a product of $X$'s along a single column (Fig.~\ref{parameterspace} (b),(c)).  We will consider a model where the check operators are randomly measured.  The fact that these checks commute with the stabilizers and logic operators can be considered as a {\em symmetry}.
 The logical operators and the Bacon-Shor stabilizers are symmetry generators. After reorganizing these operators, there is effectively an independent symmetry generator for every row and every column. Since the action of each generator is concentrated only on a lower dimensional subsystem (i.e. the spins in a single row or a single column), these are referred to as {\em subsystem symmetries}.  This designation distinguishes them from {\em global symmetries} which act on all degrees of freedom or {\em local symmetries} which involve only a single site.

 In quantum mechanics, a state is changed by the act of measuring it.  This feature is extremely powerful:  One can build a general-purpose quantum computer that solely uses a sequence of measurements to process information  \cite{measurementqc,measurementqc2}.  In this manner, random measurements introduce stochastic dynamics.  
In our first numerical experiment, we randomly measure the Bacon-Shor $\VZZ$ checks with probability $p_1$, and the  $\HXX$ checks with probability $\bar p_1 = 1-p_1$.    
We study the properties of the steady state ensemble produced by this quantum Markov chain \cite{markov}.

We characterize the steady state ensemble using spin glass order. In the limit of applying only $\HXX$ checks ($p_1=0$), each row develops a perfect X-spin glass order and similarly, only applying $\VZZ$ checks ($p_1=1$) leads to a perfect Z-spin glass order along columns. As we vary $p_1$, the ensemble displays a phase transition at $p_1=1/2$ from an independent $X$-spin glass order 
along each horizontal row for $p_1<1/2$ and an independent $Z$-spin glass order 
along each vertical column when $p_1>1/2$.  This form of order was used by Sang and Hseih to characterize the behavior of a 1D system \cite{measurementphases}, and it will be explained in detail in Sec.~\ref{subsyssymm}. We characterize the ordered phases and the 
phase transition at $p_1=1/2$.

In addition to looking at spin-glass order, we calculate the entanglement entropy between two halves of the system.  
Such entanglement based measures are analogous to thermodynamic quantities like specific heat:  They show divergences/discontinuities at phase transitions with critical exponents that encode universal features of the model and are independent of the microscopic details. 
Away from $p_1=1/2$, we find that the entanglement entropy has an {\em area law scaling} $S\sim L$, where $L$ is the linear size of our square array.  At $p_1=1/2$,  we instead see $S\sim L\ln L$,  which is referred to as 
a logarithmic violation of the area law. We explain the structure of the spin-glass order and the behavior of the entanglement entropy in terms of the subsystem symmetries of the model.

 In a recent paper \cite{Barkeshli2d}, Lavasani et al. studied a model which 
(in certain limits)
maps onto our random Bacon-Shor code.  They discovered the entanglement scaling described above, but did not explore the possibility of spin-glass order (which would be a non-local order in their model, and would not have been natural).  Another related model was  studied by Lavasani et al. \cite{lavasani} and Sriram et al. \cite{sriram}.  They considered a 2D model on a honeycomb lattice where one randomly measures non-commuting two-site checks, chosen so that each hexagonal plaquette has a conserved quantity. Hence their model contains a subsystem symmetry, but of a very different form than the Bacon-Shor code.  Their model contains a topologically ordered spin-liquid phase (with area-law entanglement entropy, $S\sim L$) and a critical phase (with entanglement entropy scaling $S\sim L\ln L$).  Neither of those phases display any spin-glass order.
 
 After characterizing the random Bacon-Shor code, we break the subsystem symmetry by allowing $\VXX$ and $\HZZ$ checks, with probability proportional to $p_2$ (Fig. \ref{parameterspace} (d)). 
In this case, the only symmetries are global symmetries, 
generated by the product of $X$ operators of all spins in the system and its $Z$-operator counterpart. With these extra $\VXX$ and $\HZZ$ check operators, we find that the spin-glass order is no longer confined to rows or columns, and that the $X$- and $Z$-spin-glass orders can coexist. At the extreme limits of applying only $\HXX , \VXX$ checks, we get a perfect global X spin glass order whereas in the limit of applying only $\HZZ, \VZZ$ checks, we get a perfect global Z spin glass order. 
One can continuously pass from the region of the phase diagram dominated by the $X$-spin-glass order to the region dominated by the $Z$-spin-glass order without encountering a phase transition. 
There is some qualitative similarity with the critical point separating the liquid and gas phases of water where one can circumvent the thermodynamic singularity by taking an appropriate path through the parameter space. 
 
 One important caveat with this analogy is that we  find discontinuities as $p_2\to 0$. An infinitesimal breaking of the subsystem symmetry leads to a discontinuous change in the spin-glass order parameters which 
describe $X$-correlations between rows or $Z$-correlations between columns.  The $X$-spin glass order in a single row (or $Z$-spin glass order in a single column) changes continuously.  In other words, an infinitesimal $p_2$ leads causes the independent spin-glass orders in individual columns and rows to merge into a global 2-dimensional spin-glass.

 Through our study, we learn that some prior heuristics have limitations.  Our model (even with the broken subsystem symmetry) features a bipartite {\em frustration graph}. That means, the check operators divide into two sets:  $\{\HXX,\VXX\}$ and $\{\HZZ,\VZZ\}$. Within each set, the operators commute with each other, but any operator from one set anti-commutes with at least one operator from the other set. In  the context of 1D models,
Ippoliti et al.~\cite{Huse} conjectured that this feature would ensure a phase transition as one varied the probabilities of measuring the various operators.  Our model instead shows a crossover, unless we impose the subsystem symmetry.

\section{Model and notation}\label{model}

\begin{figure*}
   \includegraphics[width = 18 cm]{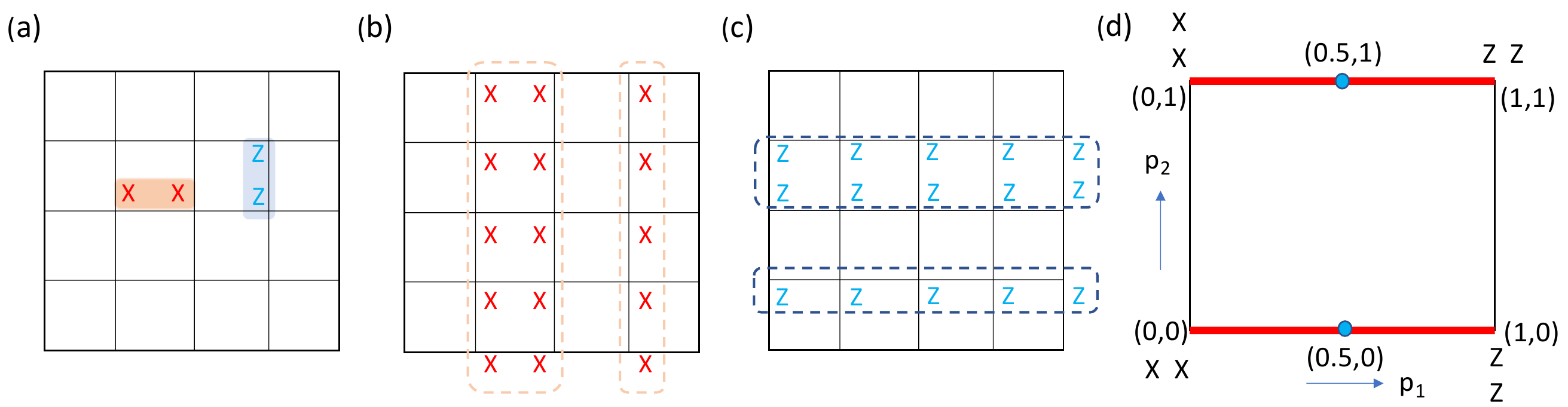}
    \caption{ The Bacon-Shor error correction model on a square lattice with qubits on the vertices (a,b,c), and the phase diagram of our measurement-only random circuit (d). (a) The operators $\HXX$ on horizontal bonds and $\VZZ$ on vertical bonds are check operators. (b,c) A product of $X$ operators along two adjacent columns and a product of $Z$'s along two adjacent rows are the Bacon-Shor code stabilizers, which can be formed from products of check operators, but commute with all checks.   The check operators also commute with the logical operators, products of $X$/$Z$ operators along a single column/row. These logical operators generate the subsystem symmetries. 
    (d) Parameter space of our model. We measure 2-qubit operators: $\HXX,\VZZ,\HZZ,\VXX$ with probabilities $\bar p_1 \bar p_2,\bar p_1 p_2,p_1 p_2,p_1\bar p_2$, where $p_1$ and $p_2$ run from 0 to 1, and $\bar p_j=1-p_j$. The solid red lines at $p_2=0,1$ illustrate the regions of parameter space having the Bacon-Shor code structure with subsystem symmetries, and the blue dots correspond to critical points.
    }
    \label{parameterspace}
\end{figure*}

We consider a square $L \times L$ grid of qubits. The qubits do not evolve under any Hamiltonian, and instead the only dynamics comes from measurements. At each time step, we measure a rantom 2-qubit check operator: $\HXX,\VZZ,\HZZ,\VXX$ with probabilites $\bar p_1 \bar p_2,\bar p_1 p_2,p_1 p_2,p_1\bar p_2$, where $p_1$ and $p_2$ run from 0 to 1, and $\bar p_j=1-p_j$.  The resulting parameter space is conveniently visualized by the square in Fig.~\ref{parameterspace}(d).  At each corner, only one type of operator is measured.  

We start with a random state in the Bacon-Shor code space and use the stabilizer formalism in Sec.~\ref{calc} to calculate how the state evolves under each measurement.
We will use $\langle O\rangle$ or $\langle\psi|O|\psi\rangle$ to denote an expectation value in a single state, and $\overline{O}$ or $\overline{\langle O\rangle}$ to denote the ensemble average. 
In our numerics, each ensemble average corresponds to time averages over 32 independent runs, each starting with a different initial state.  Each run contains $100 L^2$ measurement `time' steps.

Each qubit is labeled by indices $(i,j)$, where $i$ and $j$ are integers between 1 and $L$.  We will use Greek letters to denote the 2D coordinate, so a Pauli operator on such a qubit will be designated $X_{i,j}$ or $X_\alpha$.  Following the notation of matrices, we will consider the first index, $i$ to specify a row, while $j$ specifies a column.  This is opposite to the convention of Cartesian coordinates.

Note that we can map $p_2\to1-p_2$ by rotating the lattice by 90 degrees.  We can also map $p_1\to 1-p_1$ by simultaneously rotating the lattice and swapping $X\leftrightarrow Z$.  Thus the entire phase diagram can be constructed from one quadrant.

\subsection{Symmetries}
\label{sec::symmetries}

The system always possesses two global symmetries: The products over all sites, $\prod_{i=1}^{L} \prod_{j=1}^{L} X_{i,j}$ and $\prod_{i=1}^{L} \prod_{j=1}^{L}  Z_{i,j}$ commute with all check operators.  On  the red lines shown in Fig.~\ref{parameterspace}(d), corresponding to $p_2=0,1$, the system has additional subsystem symmetries.  When $p_2=0$, all of the check operators commute with the Bacon-Shor stabilizer generators: $\prod_{i} X_{i,j} X_{i,k}$ and $\prod_{i} Z_{j,i} Z_{k,i}$ where $j\neq k$.  On this high symmetry line, the   checks also commute with the logic operators: $\prod_{i} X_{ij}$ and $\prod_j Z_{ij}$. These operators are illustrated in Fig.~\ref{parameterspace}(b),(c). Note that the Bacon-Shor stabilizers are products of the logical operators.

The same structure is found when $p_2=1$, but with $X$ and $Z$ reversed.  For simplicity, in our numerics we always choose our initial state to be an eigenstate of the relevant symmetry operators.

\section{Calculation Technique}\label{calc}

\subsection{Stabilizer formalism}\label{stab}

We use the quantum trajectories approach in the stabilizer formalism to simulate the dynamics \cite{hybrid2d}. In the stabilizer formalism, the many-body quantum state of $L^2$ qubits is specified by $L^2$ linearly independent and mutually commuting  operators called stabilizer generators\cite{hybridcircuit1d}.

Each of these stabilizer generators is a Pauli string and the many-body quantum state is a simultaneous eigenstate of all the generators. For our calculation, we do not need to keep track of the eigenvalues, which are all $\pm 1$.  One can interpret our results as an average over all such possibilities.  The instantaneous stabilizer group formed by taking all possible products of the stabilizer generators should not be confused with the Bacon-Shor stabilizer group which describes the symmetries when $p_2=0,1$.

At each time step we randomly measure a check operator  
($\HXX,\VZZ,\HZZ$, or $\VXX$)
on a random bond on the lattice according to probabilities $p_1$ and $p_2$ defined in Sec.~\ref{model}. Based on the measurement, we update the generators. 

Each generator has the form $g=\prod_{\alpha\beta} X_{\alpha}^{n_{\alpha}}Z_{\beta}^{m_{\beta}}$ where the $n$'s and $m$'s can be 0  or 1.  Thus the  stabilizer can be stored as a length $2L^2$ binary string, $V$, containing the $n$'s and $m$'s.  In this representation two operators commute if they share an even number of non-zero entries, after the $X$ and $Z$ entries of one of them are swapped.  The generators are linearly independent if the  $L^2\times2L^2$ matrix of the $V$'s is of full rank, modulo 2.  This property can readily be checked by row reduction.

The expectation value of a Pauli operator/string $O$ can be easily calculated:  If $O$ commutes with all of the instantaneous stabilizer generators, then it must be a product of them, and $|\psi\rangle$ is an eigenstate.  Thus $\langle\psi| O|\psi\rangle=\pm1$.  Conversely, suppose $O$ anticommutes with at least one generator, $g$. (Pauli operators which do not commute instead anticommute.)  Then $\langle\psi| O|\psi\rangle=\langle\psi|g Og|\psi\rangle$ because $g |\psi\rangle=\pm |\psi\rangle$, Since $g$ and $O$ anticommute, we conclude $\langle O\rangle=0$.

When we randomly choose to measure a check operator $S$ during our dynamics,
one of two scenarios are possible: (1) If $S$ commutes with all of the generators of the instantaneous stabilizer group, then $S$ belongs to this stablizer group and the measurement does not change the state. (2) If $S$ anticommutes with generators $g_1,g_2,\cdots g_n$, the effect of the measurement can be captured by replacing the generators $g_1\to S$ and $g_j\to g_1 g_j$ for $j>1$ in the instantaneous stablizer group. The generators that initially commute with $S$ will be unchanged.

\subsection{Order Parameter}\label{sec::op}

Our model always has global symmetries generated by $\prod_\alpha X_\alpha$ and $\prod_\alpha Z_\alpha$. The single spin operators $X_\alpha$ and $Z_\beta$ anticommute with these global symmetry generators and thus the expectation values $\langle X_\alpha \rangle$ and $\langle Z_\beta \rangle$ vanish in the long time limit. 
Consequently, the lowest order correlation function which can be used to characterize the states in our ensemble are the two-point functions
$\langle X_\alpha X_\beta\rangle$ and $\langle Z_\alpha Z_\beta\rangle$.
As argued in Sec.~\ref{stab}, for a given state, these expectation values are either $0,1$, or $-1$ --  corresponding to the two spins being uncorrelated, aligned, or anti-aligned in the appropriate basis.  We calculate the ensemble average ${\cal X}_{\alpha\beta}=\overline{\langle X_\alpha X_\beta\rangle^2}$ which measures the degree of correlation between the two spins, irrespective of the sign.  ${\mathcal Z}_{\alpha\beta}=\overline{\langle Z_\alpha Z_\beta\rangle^2}$ plays the  same  role in the  $Z$-basis.  We use superscripts of $c$ or $r$ when the correlation functions are between spins on the same column or row.

A non-vanishing correlation function ${\cal X}_{\alpha\beta}$ (or ${\cal Z}_{\alpha\beta}$) at large distances indicates a spin-glass order. We refer to the spin glass order detected by ${\cal X}_{\alpha\beta}$ (${\cal Z}_{\alpha\beta}$) as the $X$-spin-glass ($Z$-spin-glass) order.

Equivalently, to detect the spin-glass orders, one can use the classic Edwards-Anderson order parameter \cite{EdwardsAnderson} which is the average of ${\cal X}_{\alpha\beta}$ or ${\cal Z}_{\alpha\beta}$ for all pairs of sites $\alpha$ and $\beta$ in the entire system (for a spin glass order with respect to a global symmetry) or within a subsystem (for a spin glass order with respect to a subsystem symmetry). Sang and Hsieh used this approach to describe the spin glass order in a 1D random circuit \cite{measurementphases}.

In principle, one could consider six different correlation functions: $XX,XY,XZ,YY,YZ,ZZ$, which can be written as matrix elements of a $3\times 3$ symmetric tensor. Of these, 
the $XX$ and $ZZ$ correlators are most relevant:  The operators $X_\alpha Z_\beta$, $X_\alpha Y_\beta$ and $Z_\alpha Y_\beta$ all anticommute with one of the global stablilizers $\prod_\alpha X_\alpha$ or $\prod_\alpha Z_\alpha$, and hence have vanishing expectation values. 
For a single realization, the $Y$-correlator $\langle Y_\alpha Y_\beta\rangle$ can be expressed as $\langle Y_\alpha Y_\beta\rangle= -\langle X_\alpha X_\beta\rangle\langle Z_\alpha Z_\beta\rangle$ due to the properties of the instantaneous stabilizer group in this model.  
By the Cauchy-Schwartz inequality, $\left(\overline{\langle Y_\alpha Y_\beta\rangle^2}\right)^2 \leq \overline{\langle X_\alpha X_\beta\rangle^2} \langle \overline{Z_\alpha Z_\beta\rangle^2} = {\cal X}_{\alpha\beta}  {\cal Z}_{\alpha\beta}$.  Thus we cannot have $Y$ spin-glass order without having both $X$ and $Z$ order.  We will therefore not report on the $Y$ correlators.


 We can always choose our instantaneous stabilizers to 
separately involve  products of Pauli $X$ operators or  products of Pauli $Z$ operators:  We call these $X$-stabilizers and $Z$-stabilizers respectively. A useful observation is that $\langle X_\alpha X_\beta\rangle^2=1$ is equivalent to the statement that 
 the only $Z$ stabilizers which pass through site $\alpha$ also pass through site $\beta$, and vice-versa.  
 This observation is a corollary of the fact that $\langle X_\alpha X_\beta\rangle^2=1$ if and only if $X_\alpha X_\beta$ commutes with every stabilizer generator.  The stabilizers which anticommute with $X_\alpha X_\beta$ are the $Z$-stabilizers which pass through only $\alpha$ or $\beta$, but not both.

When $p_2=0$, the Bacon-Shor code stabilizers (either a product of two columns of $X$ operators or two rows of $Z$ operators) are part of the stabilizer group.  Thus $\langle X_\alpha X_\beta\rangle^2=0$ unless $\alpha$ and $\beta$ lie on the same row. Similarly, $\langle Z_\alpha Z_\beta\rangle^2=0$ unless the spins lie on the same column.  (For $p_2\neq 0,1$, no such constraints are present.)  The natural interpretation of  
this row-only $X$ spin-glass order, or column-only $Z$ spin-glass order, is that each row or column forms an independent spin glass.

We find that in the presence of the subsystem symmetry ($p_2=0,1$), 
the $X$ and $Z$ spin-glass orders are mutually exclusive, When the subsystem symmetry is explicitly broken (by taking $p_2\neq0,1$), the two-types of spin glass order coexist as we will show later.

\subsection{Entanglement Entropy and Mutual Information}

In addition to the spin-glass order parameter, we characterize the states in our ensemble via their entanglement properties.  In particular, we split the system in half by a vertical cut, and calculate the resulting entanglement entropy.  We then explore how this entanglement scales with system size.

For computational simplicity, we use the second Renyi entropy,  $S_{EE} = - \ln \Tr{\rho_A^2}$, where $\rho_A$ is the density matrix formed by tracing over half the system. We calculate this entropy in the stabilizer formalism by truncating the $L^2 \times 2L^2$ matrix of the stabilizers generators to an $L^2 \times L^2$ matrix,  tracing over half of the sites in the system. We calculate the rank $R$ of this truncated matrix.  The entropy is given by, $ S_{EE}=R - L^2/2$.~\cite{randcircuit1,hybridcircuitextended}

As already argued, the entanglement entropy acts like a thermodynamic variable, and will display discontinuities or cusps at phase transitions.  It can also distinguish between certain types of states.  For example, a great deal of work has been devoted to studying phase transitions between ``area law" and ``volume law" phases in systems of dimension $d$ \cite{review,randcir1,randcir2,randcir3,randcir4,randcir5,randcir6,randcir7,randcir8,randcir9,randcir10,randcir11,randcir12,randcir13,randcir14,hybridcircuit1d,hybridcircuitextended,hybridcircuit2,Huse,Barkeshli1D,Barkeshli2d,hybrid2d}. In the area-law phases, the entanglement entropy between regions of size $L$ scales as  $L^{d-1}$, corresponding to the area of the boundary. The entanglement entropy in volume-law phases instead  scales as $L^d$.  An intuitive heuristic is that local measurements favor area-law phases,  while longer-range measurements and unitary gates favor volume law phases. Most of the previous work focuses on $d=1$. In our case, we focus on $d=2$. As we will see later, our model exhibits an interesting phase diagram without a volume-law phase.

In addition to entanglement entropy, we can also consider the mutual information between two qubits, ${\cal I}_{\alpha\beta}=S_{\alpha}+S_{\beta}-S_{\alpha\cup\beta}$.  Here $S_\alpha$ is the entanglement entropy between qubit $\alpha$ and the remainder of the system.  The quantity $S_{\alpha\cup\beta}$ is the entanglement entropy between the two spins and the rest of the system.  The mutual information captures the correlations between the two qubits.  In our system, it carries the same information as the spin correlation functions, and ${\cal I}$ can be taken as a basis independent measure of the spin-glass correlations. More precisely, our system has the global symmetries, generated by $\prod_{i,j} X_{i,j}$ and $\prod_{i,j} Z_{i,j}$, that forbid single-site operators to appear as instantaneous stabilizers. Together with the fact that the instantaneous stabilizers group is generated by $X$- and $Z$-stabilizers, the mutual information ${\cal I}_{\alpha\beta}$ between the two sites $\alpha
$ and $\beta$ can be written as ${\cal I}_{\alpha\beta} = \ln 2 \left( \langle X_\alpha X_\beta\rangle^2 +\langle Z_\alpha Z_\beta\rangle^2 \right)$.

\section{Results}

\subsection{Behavior on high symmetry line, $p_2=0$}\label{subsyssymm}

We first consider the bottom red line in Fig.~\ref{parameterspace}(d) where $p_2 =0$.  Along this line, we measure horizontal $\HXX$ checks with probability $1-p_1$ and vertical $\VZZ$ checks with probability $p_1$. 
As discussed in Sec.~\ref{sec::symmetries} these operators commute with the Bacon-Shor stabilizers and logical operators.  The resulting subsystem symmetries exclude all two-point correlation functions $\langle X_\alpha X_\beta\rangle$
along a column or 
$\langle Z_\alpha Z_\beta\rangle$
along a row.  This implies that  in the $X$-basis, the rows appear independent while in the $Z$-basis, the columns  appear independent.  These symmetry   constraints   are properties of each individual element of the ensemble and do not require ensemble averaging to observe.

\begin{figure}
\includegraphics[width = \columnwidth]{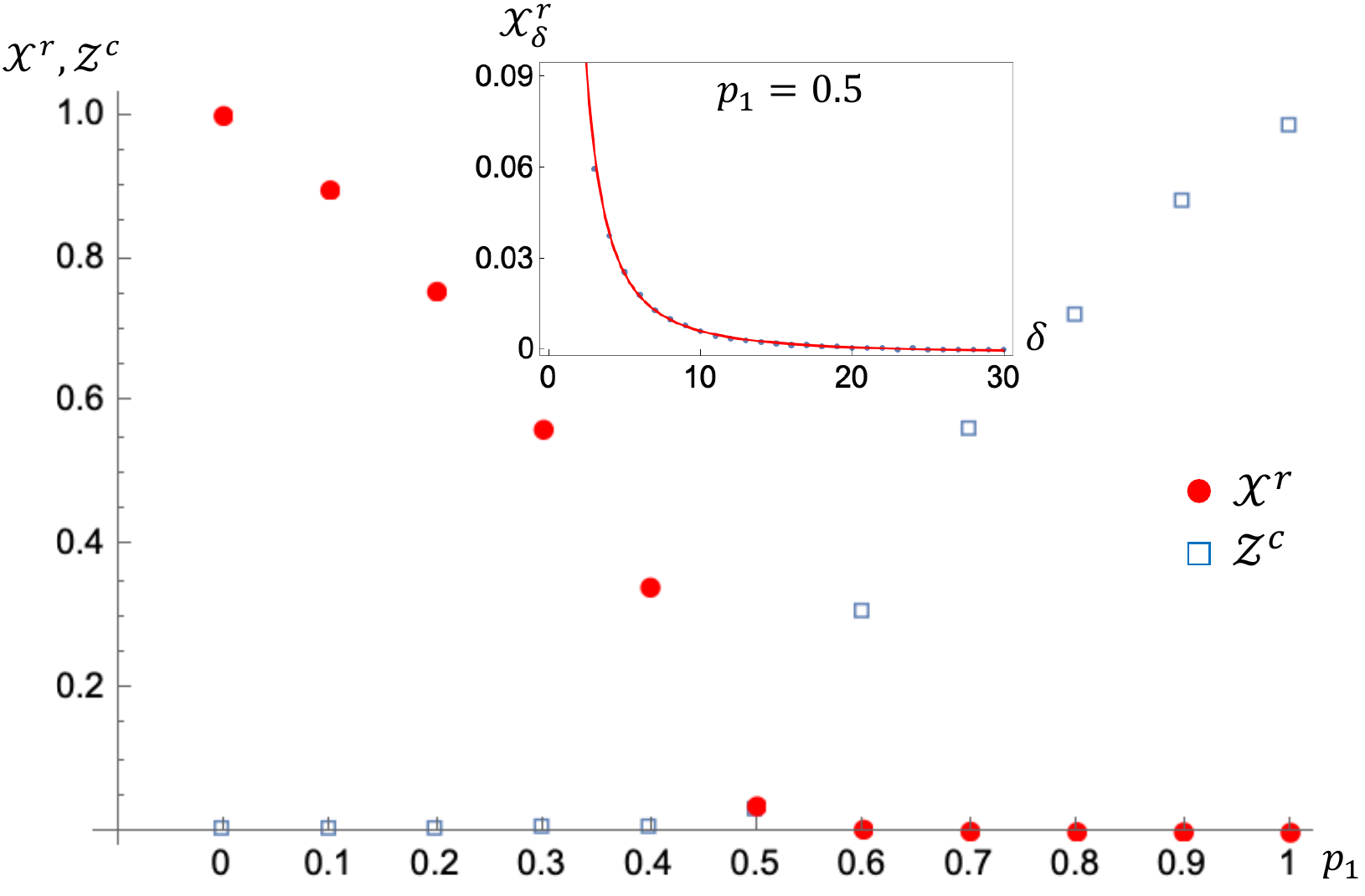}
\caption{Long-range spin-spin correlators in the $X$ basis along rows, 
 ${\cal X}^r = L^{-2}\sum_{ij}\overline{\langle X_{i,j}X_{i,j+L/2}\rangle^2}$  and in the $Z$ basis along columns, 
 ${\cal Z}^c =L^{-2}\sum_{ij} \overline{\langle Z_{i,j}Z_{i+L/2,j}\rangle^2}$, 
 as a function of probability $p_1$, along the high symmetry line $p_2=0$ for a $L\times L$ system of size $L=36$. 
The system has $X$ correlations along rows for $p_1<0.5$ and $Z$ correlations along columns for $p_1>0.5$.  The point $p_1=0.5$  is critical and there the value  of the correlator is  set by finite  size effects.  Inset:  Spatial dependence of the $X$-correlator ${\cal X}^r_\delta =L^{-2}\sum_{ij}\overline{\langle X_{i,j}X_{i,j+\delta}\rangle^2}$ at the critical point as a function of distance $\delta$ for a system of size $L=60$. The red solid line is a power law fit to the curve, $0.49 \delta^{-1.81}$.}
\label{orderparam1}
\end{figure}

We calculate the ensemble averaged long-range spin correlators in the $X$ basis along rows, 
${\cal X}^r = L^{-2}\sum_{ij}\overline{\langle X_{i,j}X_{i,j+L/2}\rangle^2}$ 
and in the $Z$ basis along columns, 
${\cal Z}^c =L^{-2}\sum_{ij} \overline{\langle Z_{i,j}Z_{i+L/2,j}\rangle^2}$ 
as we vary $p_1$ from 0 to 1 in a $36\times36$ lattice. The superscripts r and c indicate correlations along a row and column respectively. The result is shown in Fig.~\ref{orderparam1}. For $p_1<0.5$, we find ${\cal X}^r \neq 0$, indicating a spin-glass order in the $X$ basis along rows.
For $p_1>0.5$, we instead find ${\cal Z}^c \neq 0$, indicating a spin glass order in the $Z$ basis along columns. As argued below, the apparent non-zero value of ${\cal X}^r,{\cal Z}^c$ at $p_1=0.5$ is a finite size effect, and the two orders are never simultaneously present.
Due to the subsystem symmetries, the $X$-spin-glass order ($Z$-spin-glass order), if present, is separately developed on each row (column).

\begin{figure}
\includegraphics[width = \columnwidth]{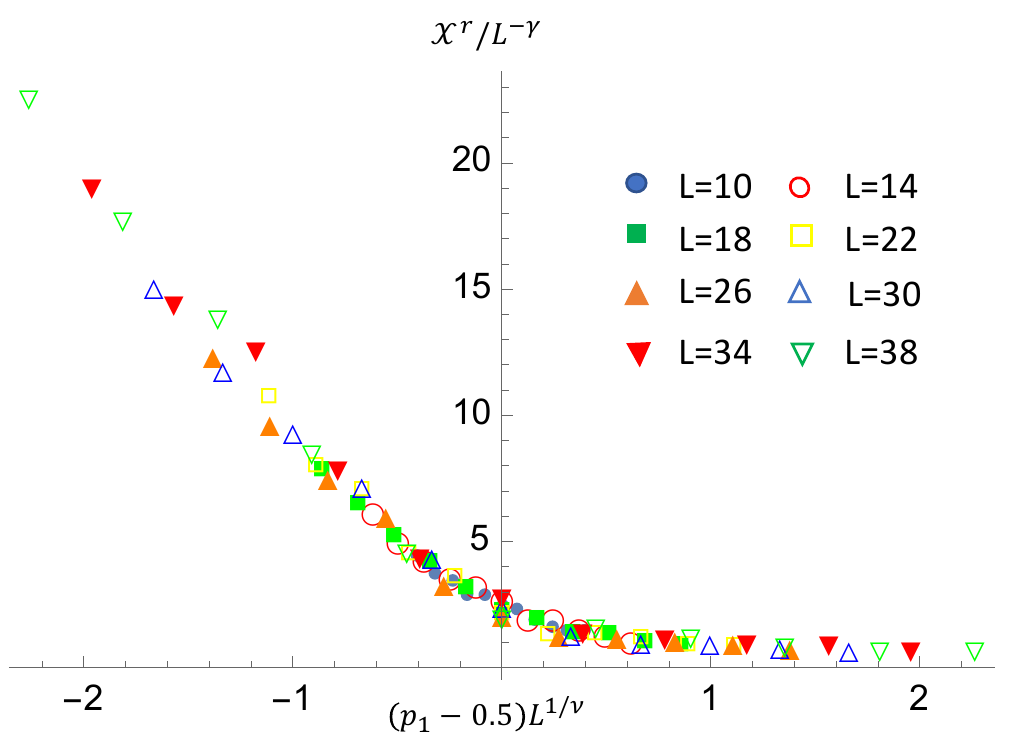}
\caption{
Scaling collapse near the critical point $p_1=0.5$, $p_2=0$, showing that
${\mathcal X}^r=L^{{- \bar \gamma}_x} g( (p_1-0.5)L^{1/\nu})$ for ${\bar \gamma}_x=1.6$ and $\nu=0.77$, where $g$ is a scaling function. }
\label{criticalcollapse}
\end{figure}

To elucidate  
the behavior at $p_1=0.5$, the inset of Fig.~\ref{orderparam1} shows 
the spatial dependence of the $X$-correlator, ${\mathcal X}^r_{\delta}= L^{-2}\sum_{i,j}\overline{\langle X_{i,j}X_{i,j+\delta}\rangle^2}$ as a function of distance $\delta$ for a larger ($L=60$) system. We see a power law decay where ${\mathcal X}^r_\delta \sim 0.49 \delta^{\gamma_x}$, with  $\gamma_x=-1.81 (5)$.  If we extrapolate to long distances, ${\mathcal X}^r_{\delta\rightarrow \infty}$ vanishes indicating the absence of $X$-spin-glass order. The $Z$-correlator ${\mathcal Z}^c_{\delta}= L^{-2}\sum_{i,j}\overline{\langle Z_{i+\delta,j}Z_{i,j}\rangle^2}$ along the columns exhibits the same scaling behavior. From this behavior, we conclude that $p_1=0.5$ is a critical point between a $Z$-spin-glass-ordered and $X$-spin-glass-ordered state. 
To further demonstrate the criticality, in  Fig.~\ref{criticalcollapse}, we show that the spin-glass order parameter obeys a scaling form
 ${\mathcal X}^r=L^{-\bar{\gamma}_x} g((p_1-0.5)L^{1/\nu})$, 
 where $g$ is a scaling function.  Following the scaling collapse procedure in Ref. \cite{hybridcircuit2},
 we find critical exponents $\bar\gamma_x = 1.6 (3)$ and $\nu = 0.77 (22)$.  Within error bars, $\gamma_x={\bar \gamma}_x$, and our value of $\nu$ agrees with the one found in Ref. \onlinecite{Barkeshli2d} by considering entropy scalings.

\begin{figure}
\includegraphics[width = \columnwidth]{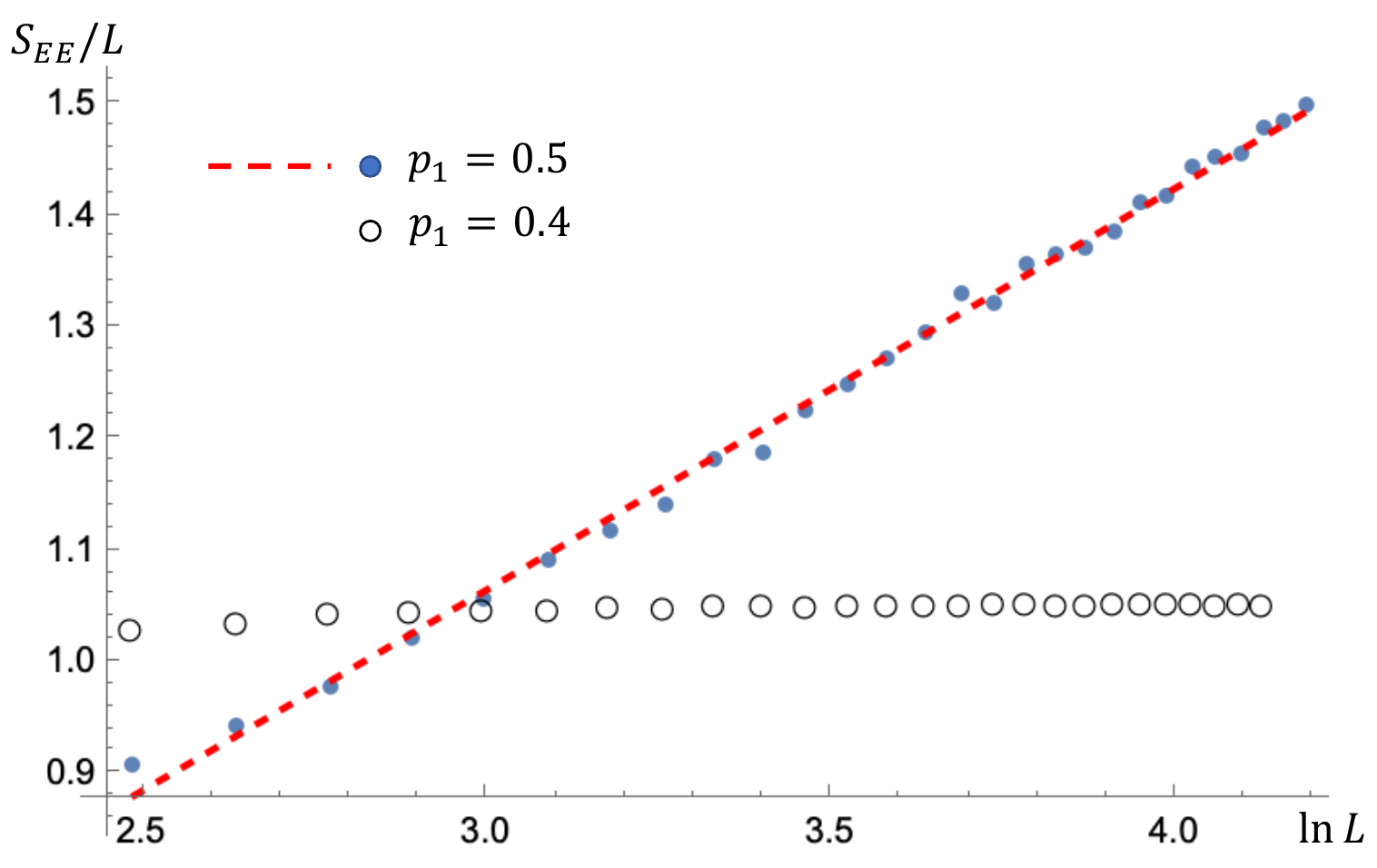}
\caption{Half cut bipartite entanglement entropy divided by system size, $S_{EE}/L$ as a function of $\ln{L}$, where $L$ is the system size of an $L \times L$ square lattice ranging from $L=12$ to $L=66$. The solid blue circles show $S_{EE}/L$ at the critical point 
$(p_1,p_2)=(0.5,0)$
. The dashed red line is a straight line fit showing logarithmic growth of $S_{EE}/L$, violating the area law at criticality. The open black circles show $S_{EE}/L$ at 
$(p_1,p_2)=(0.4,0)$
. The entropy saturates to a constant value, showing area law. 
}
\label{entropy}
\end{figure}

We also calculate the half-cut bipartite entanglement entropy, $S_{EE}$ of the system as we vary $p_1$ for different system sizes. We observe that the system obeys area-law entanglement entropy ($S_{EE} \propto L$) for all values of $p_1$ except at the critical point, $p_1=0.5$. At $p_1=0.5$, we instead find $S_{EE} \propto L \ln{L}$ as shown in Fig.~\ref{entropy}. In 1D critical points, a logarithmic violation of area-law entanglement ($S_{EE} \propto \ln L$ for a system size of L) is a common feature which is often associated with conformal symmetry. In 2D, this logarithmic violation of the area law, i.e. $S_{EE} \propto L \ln{L}$, is more unusual.
This scaling previously was found in Fermi liquids \cite{Gioev2006} and some monitored/non-unitary dynamical systems \cite{Barkeshli2d,Jian2020,hybrid2d,lnlscaling}. In our model, the scaling behavior $S_{EE} \sim L \ln{L}$ results from the subsystems symmetry, which
gives our model a 1D-like character
:
The order parameters ${\mathcal X}^r$ and ${\mathcal Z}^c$ correspond to independent ordering along each row or column.
 These rows/columns simultaneously become critical at $p_1=0.5$.  Thus it is natural to interpret the  $L\ln{L}$ entanglement scaling as being due to $\sim L$ independent 1D critical chains, each of which have $\ln L$ entanglement.
 
 It is a remarkable feature of quantum mechanics that the rows appear independent if one only looks at the $X$-component of the spins.  Conversely, the columns appear independent if one only looks at the $Z$-component.  Criticality appears simultaneously in each of these bases.
 
\subsection{Without subsystem symmetry, $p_2=0.5$}\label{nosubsyssymm}

We now consider the case where $p_2 \neq 0,1$. This breaks all the subsystem symmetries that were present when $p_2 = 0,1$. To illustrate the generic behavior, we fix $p_2=0.5$ and vary $p_1$ from 0 to 1. As will be demonstrated in Sec.~\ref{probsubsys}, other values of $p_2\neq 0,1$ show qualitatively similar behavior.  That section also discusses continuity (or the lack there-of) as $p_2\to0,1$.

 For $p_2=0.5$ measurements along horizontal and vertical bonds are equally probable.  As illustrated in Fig.~\ref{parameterspace}(d), we measure $\{\HZZ,\VZZ \}$ with probability $p_1$ and $\{\HXX,\VXX \}$ with probability $1-p_1$.  There are no constraints on the allowed form of the $X$- or $Z$-  spin glass order.  Each order can exist along rows, columns, and diagonally.

\begin{figure}
\includegraphics[width = \columnwidth]{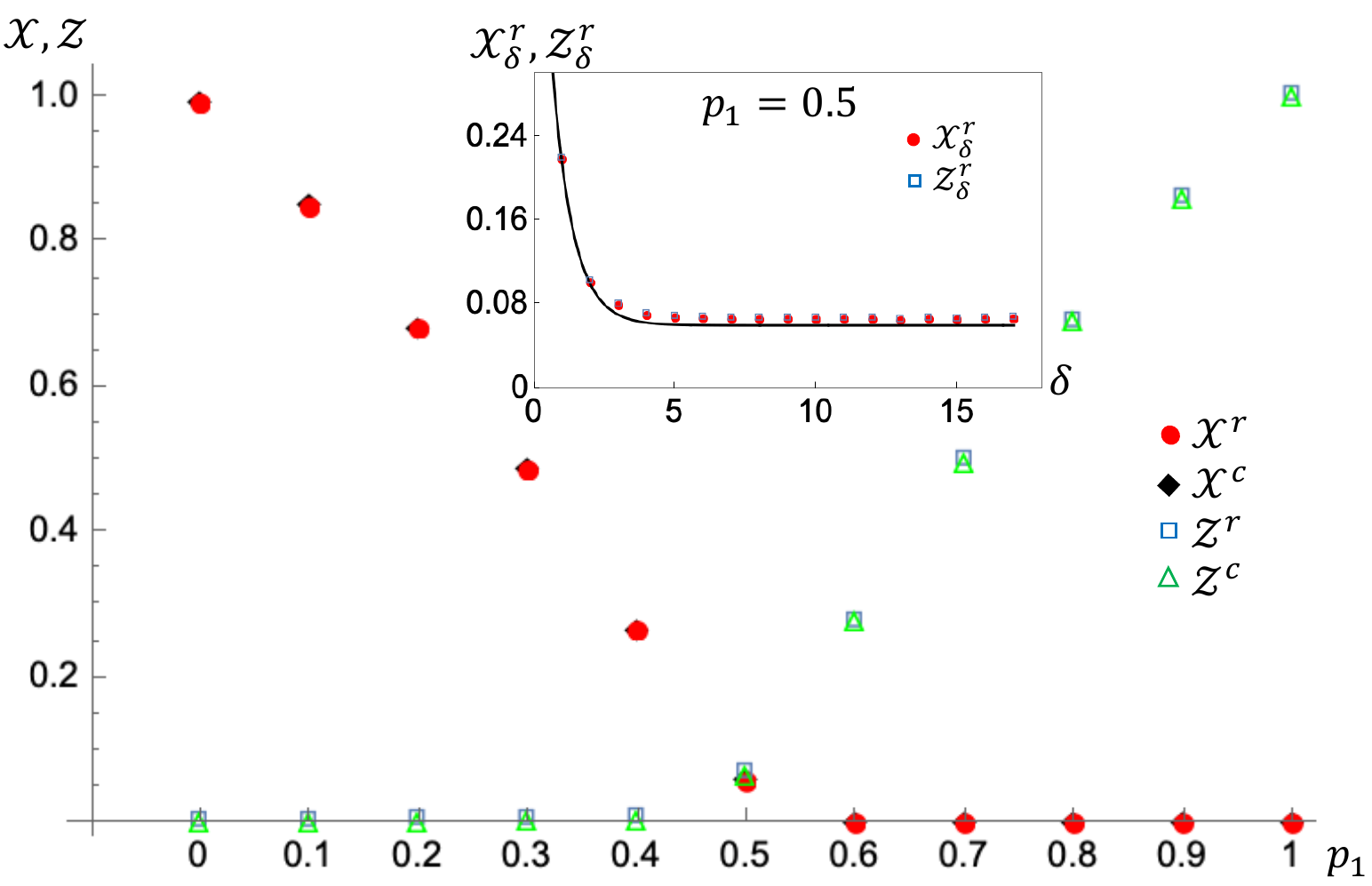}
\caption{Long-range spin-spin correlators in the $X$ and $Z$ basis along rows (${\mathcal X}^r, {\mathcal Z}^r$) and columns(${\mathcal X}^c,{\mathcal Z}^c$) as a function of probability $p_1$, with $p_2=0.5$ for an $L \times L$ system of size $L=36$.  Along this line, the correlations are isotropic: Column and row orders are of equal strength.  Near $p_1=0.5$, the $X$ and  $Z$ orders  coincide. Inset shows spatial dependence of the $X$-correlator ${\cal X}^r_\delta =L^{-2}\sum_{ij}\overline{\langle X_{i,j}X_{i,j+\delta}\rangle^2}$ and the $Z$-correlator ${\cal Z}^c_\delta =L^{-2}\sum_{ij}\overline{\langle Z_{i,j}Z_{i+\delta,j}\rangle^2}$ at the point 
$(p_1,p_2)=(0.5,0.5)$
as a function of distance $\delta$. The red solid line is an exponential fit to the curve, $0.6 e^{-1.36 \delta}+ 0.06$. The correlations fall off exponentially and saturate at a system-size independent value of 0.06.}
\label{orderparam2}
\end{figure}

In Fig.~\ref{orderparam2} we show 
the ensemble averaged long-range correlators in the $X$ and $Z$ basis along rows (${\mathcal X}^r,{\mathcal Z}^r$) and columns (${\mathcal X}^c,{\mathcal Z}^c$). Without the subsystem symmetry, the X-spin glass order and Z-spin glass order exist identically along both rows and columns. The $X$ order is peaked at $p_1=0$, while the $Z$ order peaks at $p_1=1$.  Despite the superficial similarity with Fig.~\ref{orderparam1}, the point $p_1=0.5$ is not a critical point.  Both order parameters are non-zero at $p_1=0.5$:  The $X$ order smoothly falls to zero as one increases $p_1$, and the $Z$ order behaves similarly as one decreases $p_1$.  

To demonstrate that the order parameters are finite at $(p_1,p_2)=(0.5,0.5)$, in the inset of Fig.~\ref{orderparam2}, we plot the spatial dependence of the $X$ and $Z$ correlators along rows, ${\mathcal X}^r_{\delta}= L^{-2}\sum_{i,j}\overline{\langle X_{i,j}X_{i,j+\delta}\rangle^2}$ and  ${\mathcal Z}^r_{\delta}= L^{-2}\sum_{i,j}\overline{\langle Z_{i,j}Z_{i,j+\delta}\rangle^2}$. The correlation functions fall off exponentially ($\sim 0.6e^{-1.36 \delta}$) for small $\delta$ and saturate at a finite value of ${\mathcal X}^r_{\delta\rightarrow\infty},{\mathcal Z}^r_{\delta\rightarrow\infty}\approx 0.06$ for large $\delta$. 
The long-range behavior is independent of system size as long as $L>5$.  We also calculated the $Y$ correlator, ${\mathcal Y}^r_{\delta}= L^{-2}\sum_{i,j}\overline{\langle Y_{i,j}Y_{i,j+\delta}\rangle^2}$.  We find that there is no long-range $Y$-spin-glass order.

There are three forms of averaging in calculating these correlation functions: (1) For a given state, a quantum mechanical expectation value corresponds to the average result after many measurements of identical systems.  (2) We are performing a spatial average, summing over $i$ and $j$ in calculating $\langle X_{i.j} X_{i,j+\delta}\rangle^2$.  (3) We are ensemble averaging over multiple states.  It is useful to identify the role of each of these in determining which correlation functions are zero and which are non-zero.  If we take a single state, and fix two sites, $\alpha$ and $\beta$, then (as previously explained) $\langle X_\alpha X_\beta\rangle^2$ will be 0 or 1.  Spatial averaging then gives a real number which represents the fraction of correlated sites.  We find that for large system sizes the quantities are self-averaging, meaning that the spatial average of a single realization approaches the ensemble average, as we increase system size.

This self-averaging property implies that behavior of the ensemble can be understood by studying a single typical state,
and the coexistence of $X$ and $Z$ spin-glass order at $(p_1,p_2)=(0.5,0.5)$ demonstrates that these orders can be found in each individual state.  As already explained, we find no $Y$ spin-glass order, indicating that in any given state the $X$-correlated spins are distinct from the $Z$-correlated spins.  

To further elucidate this structure, we recall that $\langle X_\alpha X_\beta\rangle\neq 0$ if and only if all $Z$-stabilizers which pass through $\alpha$ also pass through $\beta$.  Given that we are in an area-law phase, we can find a basis in which the only linearly independent long-range stabilizers are the global symmetries $\prod_\alpha Z_\alpha$ and $\prod_\alpha X_\alpha$.  The remaining generators are short-ranged, and can be classified
as either $Z$-type or $X$-type,  depending on which operators they are constructed from. Thus for well separated $\alpha$ and $\beta$, $\langle X_\alpha X_\beta\rangle\neq 0$ if and only if there are no short-range $Z$-generators passing through either of the two sites.  An analogous argument applies to $\langle Z_\alpha Z_\beta\rangle$.  

For $\langle Y_\alpha Y_\beta\rangle$ to be non-zero, there would need to be no short-range generators of any type passing through both sites. 
A simple counting argument  implies there can be at most two sites with only the global $\prod_\alpha Z_\alpha$ and $\prod_\alpha X_\alpha$ stabilizers going through them and no short range stabilizers touching them. In the thermodynamic limit, this results in no $Y$-order.

We refer to a site with no short-range $Z$-stabilizer generators as a $X$-site, and one with no short-range $X$-stabilizer generators as a $Z$-site.  All other sites have both $X$ and $Z$ short-range stabilizer generators.
In Appendix~\ref{sec::snapshot} we show typical distributions of the $X$ and $Z$ sites.  They are uniformly distributed, forming clusters whose size corresponds the length-scale over which the correlation functions decay.  The $X$-clusters and $Z$ clusters are interdigitated, and there is no global phase separation.
At $p_1=0.5,p_2=0.5$, the density of $X$-sites is $1/4$, as is the density of $Z$-sites.  This implies that the long-range correlators take the value $\langle X_\alpha X_\beta\rangle^2=(1/4)^2\approx0.06$ for well separated sites $\alpha$ and $\beta$, which is consistent with Fig.~\ref{orderparam2}.

As mentioned in the introduction, Ippoliti et al. \cite{Huse} introduced the concept of a {\em frustration graph} to  organize their thinking about random circuits.  In the frustration graph, each check operator is a node.  Two nodes are connected if they anti-commute.  Our frustration graph is bipartite, with $p_1$ controlling the weight given to the disconnected subgraphs formed respectively from measuring $\{\HXX,\VXX \}$ and $\{\HZZ,\VZZ \}$ operators.  Previous examples with a bipartite frustration graph in 1D models displayed phase transitions \cite{Huse}. Our 2D model instead  shows a smooth crossover (except in the presence of subsystem symmetries), demonstrating the lack of a direct connection between the phase diagram and frustration graph topology in higher dimensions.

The bipartite entanglement entropy obeys an area law for all values of $p_1$ when $p_2 =0.5$. In fact, the system shows area-law behavior for all values of $(p_1,p_2)$ except at the two critical points ($p_1=0.5$ and $p_2 = 0,1$) where we see an $L \ln{L}$ scaling. 

\subsection{Along $p_1 =0.25$ and $p_1 =0.5$}\label{probsubsys}

So far we have been considering fixed values of $p_2$ and varying $p_1$, making horizontal cuts in the phase diagram in Fig.~\ref{parameterspace}(d). We complete our understanding by varying $p_2$ and making vertical cuts along fixed values of $p_1=0.25, 0.5$. The parameter $p_2$ quantifies the distance from the subsystem symmetry lines at $p_2=0,1$. 

\begin{figure}
\includegraphics[width = \columnwidth]{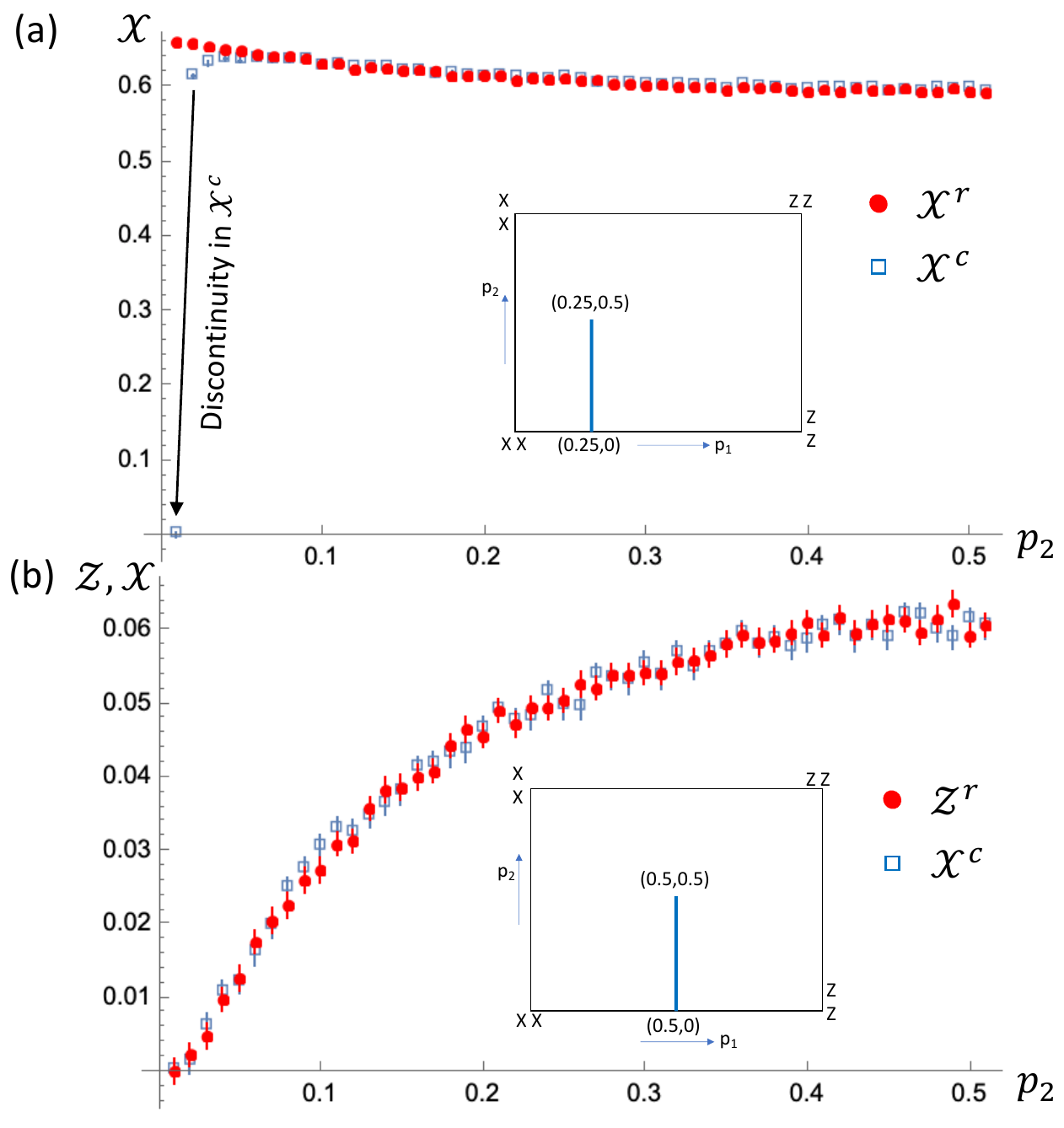}
\caption{(a) Long-range spin correlators in the $X$ basis along rows, ${\mathcal X}^r$ and columns, ${\mathcal X}^c$ as a function of $p_2$ for $p_1=0.25$ on an $L \times L$ system with $L=36$. ${\mathcal X}^r$ remains nearly constant even as $p_2 \to 0$. The order is roughly isotropic, ${\mathcal X}^c \sim {\mathcal X}^r$, when $p_2 \neq 0$, but ${\mathcal X}^c$ jumps discontinuously to 0 when $p_2 \to 0$. Inset: Blue solid line shows the region of the phase diagram probed here (varying $p_2$ at $p_1 =0.25$).
(b) Long-range spin correlators in the $X$ basis along columns, ${\mathcal X}^c$ and in $Z$ basis along rows, ${\mathcal Z}^r$ as a function of $p_2$ for $p_1=0.5$ on an $L \times L$ system with $L=36$. Both orders continuously go to zero as $p_2 \to 0$. Inset: Blue solid line shows the region of the phase diagram probed here (varying $p_2$ at $p_1 =0.5$).
Error bars represent the standard error in the mean.
}
\label{discontinuity}
\end{figure}

Figure~\ref{discontinuity} shows the long-range $X$ correlators along rows (${\mathcal X}^r$) and columns (${\mathcal X}^c$) for different values of $p_2$.  Panel (a) where $p_1=0.25$ illustrates the generic situation, away from the critical point.  Panel (b) with $p_1=0.5$ instead shows the approach to criticality. In these plots, ${\mathcal X}^r$ and ${\mathcal X}^c$ are nearly identical at non-zero $p_2$.  By considering systems of different sizes, we have established that for $p_2\neq 0$ the discrepancy between order along rows and columns is due to finite size effects and statistical noise.  

At $p_2=0$ the $X$-order along columns vanishes (${\mathcal X}^c=0$, see Sec~\ref{subsyssymm}).  
With $p_1 = 0.5$,
${\mathcal X}^c$ smoothly approaches zero as we approach the critical point by tuning $p_2\to 0$.  Conversely, for $p_1=0.25$ we find (in the limit $L\to\infty$) a discontinuity (Panel (a)). This discontinuous jump can be understood as follows. With $p_2 =0$ and $p_1<0.5$, there is effectively an independent and uncorrelated $X$-spin-glass order for every row $i$ , associated with the subsystem symmetry generated by $\prod_{j} Z_{i,j}$. Once $p_2$ becomes finite (while $p_1<0.5$), this subsystem symmetry is broken down to a global symmetry generated $\prod_{i,j} Z_{i,j}$. The existing $X$-spin-glass orders in all rows now couple to each other and merge into a 2-dimensional $X$-spin-glass. 

\section{Summary and Outlook}

Randomly measuring non-commuting observables causes a quantum system to evolve.  In the thermodynamic limit, discontinuities and singularities are found in the properties of the resulting ensembles.  These points of non-analyticity are analogous to thermodynamic phase transitions.  
As in thermodynamics, both dimensionality and symmetries are crucial to understand the universal behavior of dynamical systems undergoing both unitary evolution and measurements.
So far, the bulk of prior work in this direction has focused on the entanglement properties of 1D systems \cite{review,randcir1,randcir2,randcir3,randcir4,randcir5,randcir6,randcir7,randcir8,randcir9,randcir10,randcir11,randcir12,randcir13,randcir14,hybridcircuit1d,hybridcircuitextended,hybridcircuit2,Huse,Barkeshli1D,charge}.  Our work is part of a lively program to  enlarge this scope \cite{Barkeshli2d,lavasani,sriram,hybrid2d}.  

Our 2D model contains a number of symmetries which grow with system size (a subsystem symmetry).  This feature has important implications for its phase diagram and critical behavior.

We consider a measurement-only random circuit inspired by the Bacon-Shor error correcting code \cite{baconshor}. 
This circuit is a natural starting point, as it involves only  nearest neighbor two qubit operators, $\HXX$,$\VXX$,$\HZZ$ and $\VZZ$.
As we vary the probabilities of measuring these operators we obtain a rich phase diagram of spin-glass ordered phases  -- with features that are not seen in 1D models. 

Symmetry plays an important role in our 2D model. When we only measure $\HXX$ and $\VZZ$ ($p_2=0$), the system contains a large number of symmetries/conservation laws: the product of $X$ operators along any one column is conserved.  Similarly, the product of $Z$ operators along any row is conserved. As we vary $p_1$, the system undergoes a continuous phase transition from an $X$-basis spin glass along rows to a $Z$-basis spin glass along columns. We identify a critical point at $p_1=0.5$ where both spin glass orders vanish in the thermodynamic limit and the entanglement entropy scales as 
$S\sim L\ln L$.

We interpret this $L\ln L$ scaling behavior, often referred to as the logarithmic violation to the area law, as being due to a decoupling into 1D chains.  Remarkably, in the $Z$ basis the columns form these chains, while in the $X$ basis the rows form these chains. Criticality occurs simultaneously in these two channels. 

When we supplement the $\HXX$ and $\VZZ$ measurements with $\VXX$ and $\HZZ$ operators, the subsystem symmetries are broken. We find that the system now shows a smooth crossover between $X$ and $Z$ basis spin glass orders. The spin glass orders can coexist along both rows and columns. Without subsystem symmetries, even though the frustration graph remains bipartite, there is no criticality and the system always displays an area-law entanglement entropy. 

The system displays discontinuities as one approaches the high symmetry line at $p_2=0$.  Away from the critical point some spin glass order parameters discontinuously jump to zero as $p_2 \to 0$.  This is analogous to a first order phase transition. 

An important feature of our study is that, like Sang and Hsieh \cite{measurementphases}, we go beyond studying the entanglement properties of the quantum states, and consider the spin correlation functions.  This allows us to develop a physical understanding of the properties of the ensemble.  Similar analysis can be applied to many of the other examples in the literature \cite{Huse,Barkeshli1D,Barkeshli2d,charge,hybridcircuit1d}, though some cases show topological order, requiring the use of string operators rather than the simple two-point functions used here.

One interesting feature of our model is that, aside from the critical points, it always yields states with area-law entanglement.  Such behavior is expected when one measures sufficiently local operators.  It would be interesting to characterize the minimal set of operators which would lead to a volume-law phase in 2D.

The behaviour of 2D measurement-only random circuits is richer than their 1D counterparts.
By virtue of the larger connectivity there can exist symmetries in 2D systems without clear analogs in 1D.  Our subsystem symmetry is such an example.  In our model, this symmetry has dramatic effect, leading to discontinuities in the spin-glass order, and converting a crossover into a phase transition. It would be interesting to understand if other forms of subsystem symmetries lead to similar behaviour.

\section*{Acknowledgements}
This work was supported by 
a New Frontier Grant awarded by the  College of Arts and Sciences
at Cornell University (V.S., E.J.M.) and a faculty
startup grant at Cornell University (C.-M.J.). We thank Maissam Barkeshli and Ali Lavasani for insightful discussions.
\appendix

\section{Snapshots of spin-glass order}\label{sec::snapshot}

Here we introduce a technique to visualize the spin-glass order in a given quantum state.  
We draw a grid of colored cells, each one represents a single spin.  We color the cells according to the stabilizer generators which have  support at that location:  White cells have no short-range $Z$-stabilizers passing through them.  Black cells have no short-range $X$-stabilizers passing through them.  Gray cells have both short-range $X$ and $Z$ stabilizers.  In the language of Sec.~\ref{nosubsyssymm}, the white cells correspond to $X$-sites, and the Black cells correspond to $Z$-sites.

By construction, any two $X$-sites ($\alpha,\beta$) will be $X$-correlated,  namely $\langle X_\alpha X_\beta\rangle^2=1$.  Similarly, $Z$ sites will have $\langle Z_\alpha Z_\beta\rangle^2=1$.  Because of the area-law entanglement, these are the only long-range correlations:   In the visualizations, there are no long-range $X$ correlations involving grey cells or black cells; there are no long-range $Z$ correlations involving grey cells or white cells. There can, however, be short-range correlations of these sorts.
 
\begin{figure*}
\includegraphics[width = 18 cm]{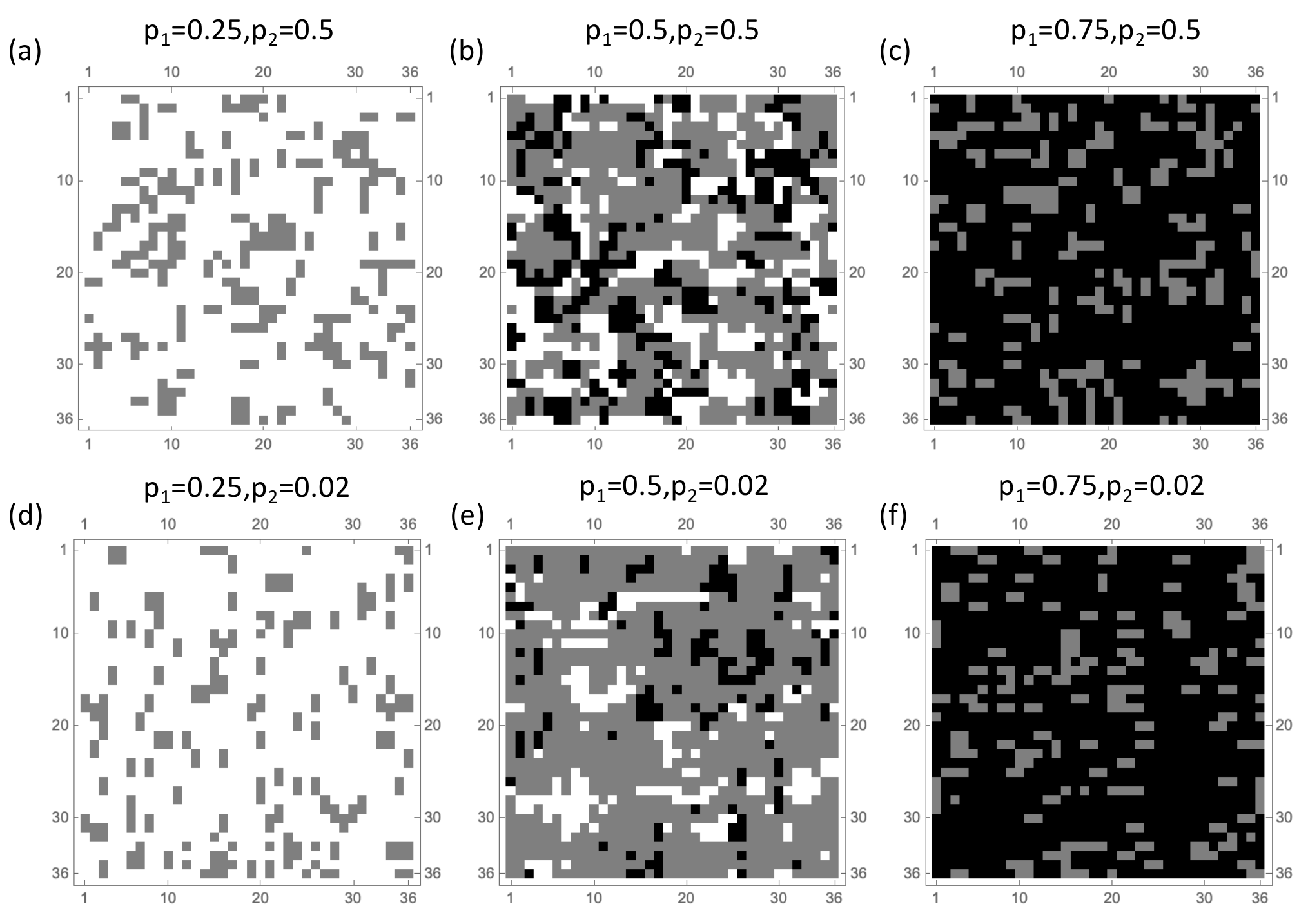}
\caption{Snapshots of typical long time quantum states for a system size $36 \times 36$ for different values of $p_1, p_2$. (a) $p_1 = 0.25, p_2 = 0.5$, (b) $p_1 = 0.5, p_2 = 0.5$, (c) $p_1 = 0.75, p_2 = 0.5$, (d) $p_1 = 0.25, p_2 = 0.02$, (e) $p_1 = 0.5, p_2 = 0.02$, (f) $p_1 = 0.75, p_2 = 0.02$.  Each colored cell represents a single spin.  White cells have no short-range $Z$-stabilizers passing through them, and are therefore $X$-correlated with one-another.  Black cells have no short-range $X$-stabilizers passing through them, and are $Z$-correlated.  Gray cells have both short-range $X$ and $Z$ stabilizers. While this coloring captures all long-range correlations, there may be additional short-range correlations.}
\label{snapshots}
\end{figure*}

We refer to each visualization as a ``snapshot".  Fig.~\ref{snapshots} shows snapshots of typical states for a variety of $p_1,p_2$.  In all cases we consider a $36\times36$ system, and run our dynamics for a large number of timesteps before producing the visualization.

As we discussed in Sec.~\ref{nosubsyssymm}, the spin glass order is self averaging  and thus a spatial average on a typical state is representative of the ensemble average. 

The top row of Fig.~\ref{snapshots} shows the snapshots when $p_2=0.5$. Fig.~\ref{snapshots}(a) shows the case where $p_1 = 0.25$. Most sites are 
$X$-correlated (white) here and spread throughout the entire system. There are no black sites, which shows that there are no long-range $Z$ correlations. Fig.~\ref{snapshots}(c) similarly shows the case when $p_1 =0.75$ where the system is dominated by $Z$-correlated sites (black) with no $X$-correlated sites (white) present. 

Fig.~\ref{snapshots}(b) shows the snapshot when $p_1=0.5$. Both white and black colored sites are randomly spread throughout the system which shows how the $X$- and $Z$-spin-glass orders coexist in the system. (Though in any given state, the sites participating in the $X$-order are distinct from the sites participating in the $Z$-order)  Roughly 1/4 of the sites are white.
For any two distant sites, the probability of them being $X$-correlated would be $(1/4)^2 \approx 0.06$. This agrees with the asymptotic behavior of the correlation function shown in the inset of 
Fig.~\ref{orderparam2}.

In the bottom row of Fig.~\ref{snapshots}, we show snapshots when $p_2 = 0.02$. These points are close to the subsystem symmetry line but are still area-law entangled. Fig.~\ref{snapshots}(d) and Fig.~\ref{snapshots}(f) show the states with dominant $X$ correlations and $Z$ correlations respectively. The number of white and black sites are nearly identical to what is seen when $p_2 = 0.5$ (Fig.~\ref{snapshots}(a),(c)) confirming the results of Fig.~\ref{discontinuity}(a) where the spin glass orders are only weakly dependent on $p_2$. However, 
we see that the uncorrelated gray clusters display interesting asymmetries:  They are more horizontal for $p_1=0.25$   and more vertical when $p_1=0.75$. 
This elongation is an intuitive consequence of the different probabilities of measuring horizontal or vertical check operators.
The asymmetry can also be seen in the short-range spin correlation functions -- though the long-range correlation functions are rotationaly symmetric.  


Fig.~\ref{snapshots}(e) shows the case where $p_1 = 0.5$. While we see both $X$-correlated (white) and $Z$-correlated (black) sites randomly spread throughout the system, the image is dominated by the uncorrelated gray sites. This is a visual indicator of the feature of Fig.~\ref{discontinuity}(b) where the $X$- and $Z$-spin-glass orders continuously fall to zero as $p_2 \to 0$.  As we make $p_2$ smaller, we will continue to see more and more uncorrelated sites.

\bibliography{main.bib}

\end{document}